\documentclass{article}
\usepackage{authblk}
\usepackage{graphicx}

\begin{document}

\title{Giant chiro-optical response in high harmonic generation}
	
\author[1,2]{David Ayuso\thanks{david.ayuso@mbi-berlin.de}}
\author[2,3]{Andres F. Ordonez}
\author[4]{Piero Decleva}
\author[1,2,5]{Misha Ivanov}
\author[2,6]{Olga Smirnova\thanks{olga.smirnova@mbi-berlin.de}}
\affil[1]{Department of Physics, Imperial College London, SW7 2AZ London, United Kingdom}
\affil[2]{Max-Born-Institut, Max-Born-Str. 2A, 12489 Berlin, Germany}
\affil[3]{ICFO, The Barcelona Institute of Science and Technology, Av. Carl Friedrich Gauss 3, 08860 Castelldefels (Barcelona), Spain}
\affil[4]{Dipartimento di Scienze Chimiche e Farmaceutiche, Universit\`a degli Studi di Trieste, Via L. Giorgieri 1, 134127 Trieste, Italy}
\affil[5]{Institute f\"ur Physik, Humboldt-Universität zu Berlin, Newtonstr. 15, 12489 Berlin, Germany}
\affil[6]{Technische Universit\"at Berlin, Hardenbergstr. 33A, 10623 Berlin, Germany}
\date{}
\maketitle

\begin{abstract}
High harmonic generation (HHG) records the ultrafast electronic response of matter to light, encoding key properties of the interrogated quantum system, such as chirality.
The first implementation of chiral HHG [Cireasa et al \emph{Nat. Phys.} {\bfseries 11}, 654 (2015)] relied on the weak electronic response of a medium of randomly oriented chiral molecules to the magnetic component of an elliptically polarized wave, yielding relatively weak chiro-optical signals.
Here we show that elliptically polarized light can drive a strong chiral response in chiral molecules via purely electric-dipole interactions -- the magnetic component of the wave does not participate at all.
This giant chiro-optical response, which remains hidden in standard HHG experiments, can be mapped into the macroscopic far-field signal using a non-collinear configuration, creating new opportunities for bringing the electric-dipole ``revolution'' to HHG.
\end{abstract}

Chirality is a ubiquitous property of matter, from its elementary constituents to molecules, solids \cite{Bode2007} and biological species \cite{Jordan1927}.
Chiral molecules appear in pairs of enantiomers, two virtually identical versions in which their nuclear arrangements present non-superimposable mirror images of each other.
Separated only by a mirror, left- and right-handed objects are readily distinguished in everyday life, but not in the micro-world.
Standard methods of chiral recognition based on interrogating chiral molecules with light rely on their electronic response to both the electric and magnetic components of the light field 	\cite{book_Berova2013}.
The weakness of magnetic interactions leads to weak chiro-optical signals and a justified impression that chiral discrimination is difficult, especially on ultrafast time scales.
The electric-dipole ``revolution'' \cite{Ayuso2021_perspectives} in chiral discrimination is steadily changing this perception.

Several methods \cite{Ritchie1976PRA,Powis2000JCP,Bowering2001PRL,Garcia2013NatComm,Patterson2013Nat,Patterson2013PRL,Eibenberger2017PRL,Fischer2000PRL,Belkin2001PRL,Fischer2002CPL,Beaulieu2018NatPhys} working under a unified paradigm \cite{Ordonez2018PRA} address a striking variety of physical processes, from molecular photoionization with XUV light \cite{Ritchie1976PRA,Powis2000JCP,Bowering2001PRL,Garcia2013NatComm} to microwave excitation of molecular rotations \cite{Patterson2013Nat,Patterson2013PRL,Eibenberger2017PRL} to sum-frequency generation in chiral liquids driven by infrared (IR) and visible fields \cite{Fischer2000PRL,Belkin2001PRL,Fischer2002CPL} to vibronic excitations in molecules triggered by UV light \cite{Beaulieu2018NatPhys}.
All these methods detect an enantio-sensitive and dichroic vectorial observable \cite{Ordonez2018PRA}: photoelectron current or induced polarization. 

Photoelectron circular dichroism (PECD) is the first method that heralded the electric-dipole revolution \cite{Ritchie1976PRA,Powis2000JCP,Bowering2001PRL,Garcia2013NatComm}. 
The original setup relies on one-photon ionization of randomly oriented chiral molecules by circularly polarized light, and detects asymmetry in photo-electron emission along the light propagation direction, the so-called forward-backward asymmetry.
The photo-electron current resulting from this asymmetry can involve a few tens of percent of the total number of emitted photo-electrons, and is absent in randomly oriented ensembles of achiral molecules.
That is, the chiral response in PECD reaches a few tens of percent, numbers previously unheard of in molecular chiral discrimination.
PECD has been more recently extended to the multi-photon regime \cite{Lux2012Angew,Lehmann2013JCP,Kastner2016CPC}, where two-color implementations are possible \cite{Demekhin2018PRL,Goetz2019PRL,Rozen2019PRX}.

High harmonic generation (HHG) captures the electronic response of atoms, molecules and solids to intense laser fields, with sub-femtosecond temporal resolution.
In turn, this response encodes key properties of the interrogated quantum system, exposing such intriguing phenomena as topological phases of matter \cite{Silva2019NatPhot,Bai2020NatPhot}, magnetism \cite{LaOVorakiat2019PRL} and chirality \cite{Cireasa2015NatPhys,Smirnova2015JPB,Ayuso2018JPB,Ayuso2018JPB_model,Harada2018PRA,Baykusheva2018PRX,Baykusheva2019PNAS,Neufeld2019PRX,Ayuso2019NatPhot,Ayuso2021NatComm,Ayuso2021_optica}.
This highly nonlinear process, which converts intense radiation, usually in the IR domain, into high-energy photons, can be understood semi-classically as a sequence of three steps \cite{Smirnova2014_book}:
(i) strong-field ionization,
(ii) propagation of the liberated electron in the continuum,
and (iii) electron-ionic core recombination resulting in the emission of harmonic light.

\begin{figure}[htbp]
\centering\includegraphics[width=\linewidth, keepaspectratio=true]{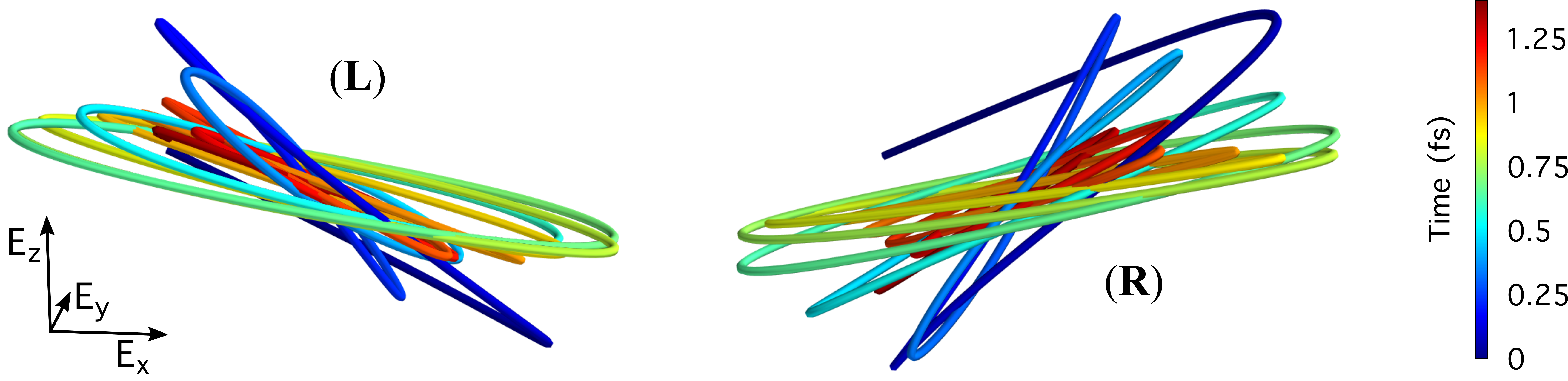}
\caption{\textbf{Time evolution of a macroscopic chiral dipole.}
Time-dependent polarization of a medium of randomly oriented propylene oxide molecules driven by a laser field with intensity $I_0=5\cdot10^{13}$ W cm$^{-2}$, wavelength $\lambda=1770$ nm and 5$\%$ of ellipticity (see \cite{Ayuso2019NatPhot} for details of the calculations).}
\label{fig_dipole}
\end{figure}

The third step of HHG, photo-recombination, is the inverse process of photo-ionization.
Thus, one would expect that HHG from randomly oriented chiral molecules should faithfully reproduce the chiral response of PECD.
This analogy also suggests that, just like the enantio-sensitive current in PECD, the coherent dipole associated with photo-recombination should have a component that oscillates in the direction of light propagation that is exactly out of phase in opposite enantiomers, reflecting the sign flip of forward-backward asymmetry in PECD.
While this perspective is correct (see Fig. \ref{fig_dipole}), it is also disheartening.
The chiral dipole component that oscillates along the propagation direction of the incident field radiates orthogonal to this direction, escaping observation in a standard high harmonic setup. 
Indeed, the HHG signal benefits from the coherent constructive addition of light bursts generated by each individual molecule, which is only possible for the harmonic light co-propagating with the driving field.
This is why in the recent work on HHG from chiral media \cite{Cireasa2015NatPhys,Smirnova2015JPB,Ayuso2018JPB,Ayuso2018JPB_model,Harada2018PRA,Baykusheva2018PRX,Baykusheva2019PNAS}  
the enantio-sensitive signal arose via a mechanism involving the magnetic field component of the laser pulse.
A similar limitation plagues wave-mixing processes in the optical range.

To overcome this limitation, Fischer and co-workers \cite{Fischer2000PRL} used a non-collinear geometry of driving fields to obtain sum-frequency generation from chiral liquids, but observed a vanishingly small chiral signal due to the non-resonant interaction of the infrared driving field with the medium.
The situation changes dramatically for high-order harmonics, with photon energies above the ionization potential \cite{Neufeld2019PRX,Ayuso2019NatPhot,Ayuso2021NatComm,Ayuso2021_optica}.
We have recently proposed several strategies to image molecular chirality on ultrafast time-scales via HHG spectroscopy, either using two-colour fields in non-collinear geometries  \cite{Neufeld2019PRX,Ayuso2019NatPhot,Ayuso2021NatComm},
or a single beam carrying one central frequency, but confined in time and space \cite{Ayuso2021_optica}, i.e. an ultrashort and tightly focused pulse.
These setups enable highly efficient and ultrafast imaging of chiral dynamics in gas-phase samples and liquid microjets.
However, their application to thick and optically dense media, e.g. a liquid cell, is more challenging, because the two-colour phase delay \cite{Neufeld2019PRX,Ayuso2019NatPhot,Ayuso2021NatComm} or the carrier-envelope phase \cite{Ayuso2021_optica} are required to remain constant as the driving field propagates through the sample in order to maximize the macroscopic chiro-optical signal.

Is it possible to simplify our previous proposals \cite{Neufeld2019PRX,Ayuso2019NatPhot,Ayuso2021NatComm,Ayuso2021_optica}, so that they could be more easily extended to thick liquid media, or even amorphous chiral solids?
What would be the ``price to pay'' for such a simplification?
Here we provide answers to these questions by introducing an all-optical method for ultrafast and highly efficient imaging of chiral matter which does not require multi-colour or ultrashort laser pulses, and therefore could be more easily applied to liquids, the natural medium of biology.

Our key results are as follows.
First, we show that the highly nonlinear response of randomly oriented chiral molecules to elliptically polarized mid-IR fields has a strong component that is orthogonal to the polarization plane of the driver, which is exactly out of phase in opposite molecular enantiomers, and that could not be observed with the single-beam configuration used in \cite{Cireasa2015NatPhys}.
Second, we present a non-collinear optical setup that allows us to map this microscopic chiral response into a giant macroscopic chiro-optical signal, in the far field, by ensuring constructive addition of the emission by individual molecules.
We also show how this setup separates the chiral and non-chiral harmonic emission in frequency, polarization and space, sending the chiral signal in a direction different from that of the non-chiral signal, enabling background-free detection.
Last but not least, we propose several strategies to measure this enantio-sensitive phase, which encodes the handedness of the chiral medium.

The first chiral HHG experiments \cite{Cireasa2015NatPhys} recorded the intensity of emission of the chiral molecules propylene oxide and fenchone in a standard single-color, single-beam geometry, using an elliptically polarized incident laser pulse.
In this case, the chiral macroscopic high harmonic response involves both the magnetic- and electric-field components of the laser pulse, and leads to relatively weak enantio-sensitivity due to the weak nature of magnetic-dipole transitions in mid-IR laser fields.

Let us consider an intense, elliptically polarized, mid-IR laser field, as in \cite{Cireasa2015NatPhys}, that propagates along $z$ and is elliptically polarized in the $xy$ plane.
Fig. \ref{fig_dipole} shows that, for randomly oriented propylene oxide molecules and $5\%$ of ellipticity, the high harmonic dipole acquires a large polarization component along $z$, i.e. orthogonal to the plane of polarization of the driving field and parallel to its propagation direction.
It is induced solely by electric-dipole interactions, it arises only due to molecular chirality, and it is out of phase in opposite molecular enantiomers.
However, this strong chiral response does not contribute to the macroscopic signal if driving laser field propagates along $z$, i.e. orthgonal to its polarization plane, as in the single-beam configuration used in \cite{Cireasa2015NatPhys}. 

\begin{figure}[htbp]
\centering\includegraphics[width=\linewidth, keepaspectratio=true]{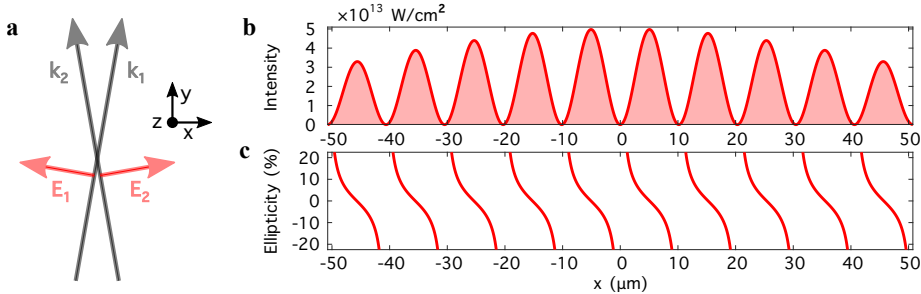}
\caption{\textbf{Proposed optical setup.}
\textbf{a,} Schematic representation of the non-collinear setup: two non-collinear, linearly polarized beams with the propagation vectors $\mathbf{k}_1$ and $\mathbf{k}_2$ and co-planar electric fields $\mathbf{E}_1$ and $\mathbf{E}_2$ induce ``forward'' ellipticity in the overlap region.
\textbf{b,} Intensity and forward ellipticity $\varepsilon_f=E_y/E_x$ in the focus along the $x$ direction.
The intensity in each beam is $1.3\cdot10^{13}$ W$\cdot$cm$^{-2}$ and they form angles of $\pm5^{\circ}$ with respect to the $y$ axis.}
\label{fig_setup}
\end{figure}

This problem is solved in the setup shown in Fig. \ref{fig_setup}.
It involves two non-collinear beams (Fig. \ref{fig_setup}a) with wavevectors $\mathbf{k}_1$, $\mathbf{k}_2$ propagating in the $xy$ plane and forming angles $\pm \alpha$ with respect to the $y$ axis.
The beams are linearly polarized in the $xy$ plane.
Thanks to non-zero $\alpha$, the total field can be elliptically polarized in the plane of propagation, with the main component along the $x$ axis and the minor component along the $y$ axis.
The regions with sufficiently high total intensity (Fig. \ref{fig_setup}b) and moderate ``forward'' ellipticity (Fig. \ref{fig_setup}c), or transverse spin \cite{Bliokh2015}, form a periodic structure across the focus where high harmonics are efficiently generated. 
In contrast to the single-beam configuration, now the chiral ($z$-polarized) component of the induced dipole (Fig. \ref{fig_dipole}) is ideally suited for generating a macroscopic phase-matched response in the far field:
oscillations along $z$ lead to harmonic emission in the $xy$ propagation plane of the incident fields.

\begin{figure}[htbp]
\centering\includegraphics[width=\linewidth, keepaspectratio=true]{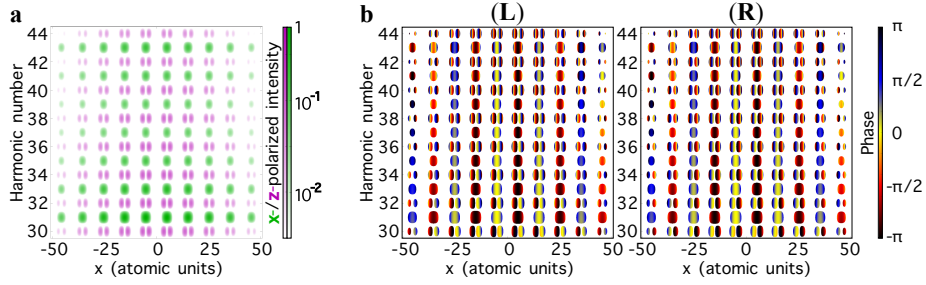}
\caption{\textbf{HHG in the near field.}
\textbf{a,} Intensity of HHG driven by the non-collinear setup presented in Fig. \ref{fig_setup} in a medium of randomly oriented propylene oxide molecules (either left- or right-handed), as a function of the transverse coordinate $x$.
\textbf{b,c,} Relative phase of the highly nonlinear response of the left-handed (\textbf{b}) and right-handed (\textbf{c}) enantiomers.
Note that the phase of the nonlinear response of the opposite enantiomers is identical for odd harmonic frequencies and opposite (i.e. with a $\pi$ difference) for even harmonic frequencies.  
}
\label{fig_near}
\end{figure}

We have calculated the highly nonlinear response of randomly oriented propylene oxide molecules to the optical setup presented in Fig. \ref{fig_setup} using the semi-analytical method described in \cite{Ayuso2019NatPhot}.
The periodic structure of the driving field along $x$ (Fig. \ref{fig_setup}) forms amplitude and phase gratings in the generated nonlinear response, i.e. a macroscopic space- and time-periodic structure, see Fig. \ref{fig_near}.
It oscillates on the attosecond time-scale, with the chiral response being out of phase for the two opposite enantiomers.

The optical response of the chiral medium to the setup of Fig. \ref{fig_setup} is not enantio-sensitive in intensity, see Fig. \ref{fig_near}a, because the driving field has mirror symmetry, e.g. with respect to the $xy$ plane, and therefore it is achiral.
Note also that this highly nonlinear response contains even and odd harmonics, which are polarized orthogonal to each other.
The achiral response appears at odd harmonic frequencies, is polarized along $x$, and is essentially driven by the strong-field ($x$-polarized) component of the laser field, as in standard HHG in achiral media.
It is identical in left- and right-handed molecules, see Fig. \ref{fig_near}.
The chiral response appears at even harmonic frequencies and is polarized along $z$.
It is produced via a ``corkscrew-like'' mechanism: the chiral medium turns the elliptical (in-plane) polarization of the light wave into out-of-plane chiral electron motion.
It has the same intensity (Fig. \ref{fig_near}a) and opposite phase (Figs. \ref{fig_near}b) in opposite molecular enantiomers.

\begin{figure}
\centering\includegraphics[width=\linewidth, keepaspectratio=true]{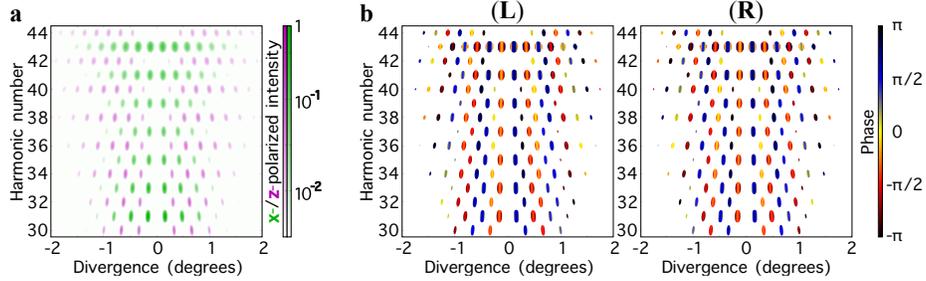}
\caption{\textbf{HHG in the far field.}
\textbf{a,} Intensity of HHG emitted from an enantio-pure sample of propylene oxide (either left- or right-handed) using the laser configuration of Fig. \ref{fig_setup} (see Fig. \ref{fig_near} for the near-field results).
\textbf{b,c,} Relative phase of the harmonic light emitted from the left-handed (\textbf{b}) and right-handed (\textbf{c}) enantiomers.
Note that phase of the light emitted by the opposite enantiomers is identical for odd harmonic frequencies and opposite (i.e. with a $\pi$ difference) for even harmonic frequencies.  
}
\label{fig_far}
\end{figure}

Fig. \ref{fig_far} shows how the different intensity and phase gratings of the chiral and achiral polarization components in the near field (Fig. \ref{fig_near}) translate into the far field: even and odd harmonics are emitted in different directions.
Thus, the even harmonics constitute a background-free measurement of molecular chirality, separated not only in frequency and polarization, but also in space.
The origin of the different emission directions can also be understood in terms of conservation of linear momentum, which dictates that, for small $\alpha$, a photon with frequency $N\omega$ can only be emitted at angles $\simeq \frac{\Delta N}{N} \alpha$,
where $\Delta N$ is the difference between the number of photons absorbed from the first and second beams.
Even and odd harmonics are emitted at different angles because $\Delta N$ can only take values that have the same parity as $N$.

Note that, like in the near-field response (Fig. \ref{fig_near}), the intensity of harmonic emission in the far-field, shown in Fig. \ref{fig_far}a, is not enantio-sensitive because the driving field is achiral.
Also like in the near-field response, the handedness of the chiral medium is encoded in the phase of the chiral response: opposite molecular enantiomers emit light at even harmonic frequencies with opposite phase, see Fig. \ref{fig_far}b.

\begin{figure}[htbp]
\centering\includegraphics[width=\linewidth, keepaspectratio=true]{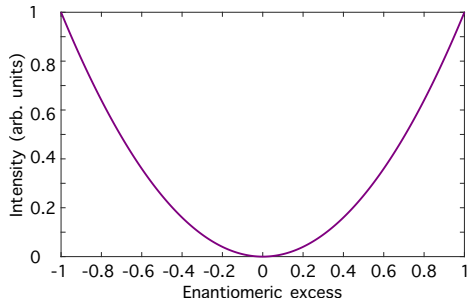}
\caption{\textbf{Working with mixtures of opposite molecular enantiomers.}
Ratio between the intensity of consecutive harmonics $I_{2N}/I_{2N+1}$ as a function of the enantiomeric excess, $ee=\frac{C_R-C_L}{C_R+C_L}$.}
\label{fig_ee}
\end{figure}

In a mixture of opposite enantiomers, the chiral contributions from left- and right-handed molecules to even HHG interfere destructively, creating opportunities for enantiomeric quantification.
The relative concentration of opposite enantiomers in a mixture is usually quantified via the enantiomeric excess, $ee=\frac{C_R-C_L}{C_R+C_L}$, where $C_R$ and $C_L$ are the concentrations of right- and left-handed molecules.
Because the intensity odd HHG is not sensitive to the medium's handedness, in our setup the intensity ratio between consecutive harmonics $I_{2N}/I_{2N+1}$ is proportional to $ee^2$, for fixed $C_R+C_L$.
Thus, the absolute value of $ee$ can be retrieved from intensity measurements, see Fig. \ref{fig_ee}.
Note that $C_R+C_L$, i.e. the total concentration of molecules in the problem sample, can be retrieved via optical or chemical methods that are not sensitive to chirality.
The quadratic dependence of the intensity of the chiro-optical signal with $ee$ is similar to that found in sum-frequency generation in the perturbative regime \cite{Fischer2002CPL}. 
The sign of $ee$, i.e. the handedness of the most abundant enantiomer in the mixture, remains hidden in the phase of the even harmonics.

We can retrieve the phase of the even harmonics, which encodes the medium's handedness, by making them interfere with a reference signal, either in the near field or in the far field.
Far-field interference can be achieved by adding an optical path to the setup that contains a BBO crystal, which would create an orthogonally polarized second harmonic of the fundamental field.
One could then create the reference signal via (achiral) HHG with the doubled-frequency field, e.g. from an atomic gas, and make it interfere with the chiral signal emitted by the chiral medium.

Alternatively, one can use the chiral medium to create the achiral reference signal in situ, making the near-field intensity already enantio-sensitive. 
This can be achieved by adding to the setup of Fig. \ref{fig_setup}, in both beams, a weak second harmonic field that is orthogonally polarized, also using a BBO crystal, creating \emph{synthetic} chiral light \cite{Ayuso2019NatPhot}.
This second harmonic field would gently push the electron out of the $xy$ plane during its trajectory in the continuum (second step of HHG) in a controlled way, and would make it recombine with a small velocity component along $z$, inducing a polarization component in this direction.
This achiral polarization component would interfere with the chiral polarization component, yielding enantio-sensitive intensity.
Synthetic chiral light that is \emph{locally} and \emph{globally} chiral \cite{Ayuso2019NatPhot} enables the ultimate efficiency limit in chiral discrimination at the level of total signal intensities, quenching it in one enantiomer while maximizing it in its mirror twin.
However, it requires having (at least) two colours in the driving field, with controlled phase delays \cite{Ayuso2019NatPhot}.

We can also generate a reference signal that will interfere with the chiral signal of the problem sample already in the interaction region (i.e. in the near field) without adding extra colours to the driver, by mixing the problem sample of unknown handedness $h_0=\pm1$, with a reference chiral sample of known handedness $h_{r}=\pm1$.
The contribution to even HHG from the problem sample is proportional to $h_0 C_0 P_0$, where $C_0$ is the concentration of molecules of unknown handedness and $P_0$ is the single-molecule contribution to HHG from a right-handed molecule at an even harmonic frequency.
Likewise, the contribution from the reference chiral substance to even HHG is $h_r C_r P_r$.
The total intensity of even HHG becomes proportional to $|h_0 C_0 P_0 + h_r C_r P_r|^2$, and thus sensitive to the handedness of the problem sample $h_0$.
The situation is particularly favorable if we choose as the reference chiral molecule (with known handedness $h_r$) the chiral molecule of the problem sample, because then $P_r=P_0$, and make $C_r=C_0$.
This will maximize even HHG when the problem and reference samples have the same handedness ($h_0=h_r$), and fully quench it they are of opposite chirality ($h_0=-h_r$).

The possibility of driving strong chiro-optical signals via HHG using a non-collinear setup that carries only one single frequency creates new opportunities for efficient chiral recognition and ultrafast imaging of molecular chirality.
The mechanisms responsible for this giant chiro-optical response are general to chiral media in the gas, liquid and condensed phases.

The proposed all-optical method can be seen as a simplified version of our previous proposals \cite{Neufeld2019PRX,Ayuso2019NatPhot,Ayuso2021NatComm,Ayuso2021_optica} that can be more easily implemented in a lab.
Of course, this simplification comes at a price: the emitted harmonic light is no longer enantio-sensitive in intensity (unlike in \cite{Ayuso2019NatPhot}), is not unidirectionally bent (unlike in \cite{Ayuso2021NatComm}), and it does not have enantio-sensitive elliptical polarization (unlike in \cite{Neufeld2019PRX,Ayuso2021_optica}).
Yet, it allows us to discriminate between chiral and achiral media with high efficiency and ultrafast temporal resolution, and even to retrieve the absolute value of the enantiomeric excess in mixtures of left- and right-handed molecules of unknown concentrations.
It also offers a great advantage with respect to our previous proposals \cite{Neufeld2019PRX,Ayuso2019NatPhot,Ayuso2021NatComm,Ayuso2021_optica}: here, because the driving field carries a well-defined single frequency, there are no two-colour phase delays (unlike in \cite{Neufeld2019PRX,Ayuso2019NatPhot,Ayuso2021NatComm}) or carrier-envelope phases (unlike in \cite{Ayuso2021_optica}) that need to be controlled, which simplifies the experimental realization and brings new opportunities for measuring giant chiro-optical signals in the liquid phase, the natural medium of biological molecules, or in amorphous chiral solids. 
Because of the ultrafast temporal resolution enabled by HHG spectroscopy, this relatively simple all-optical setup seems to be ideally suited to monitor chiral-to-achiral phase transitions in matter, e.g. an ultrafast chemical reaction where a chiral molecule loses its handedness and becomes achiral.

\section*{Acknowledgments}
Royal Society (URF$\backslash$R1$\backslash$201333).
Engineering and Physical Sciences Research Council (MURI-MIR EP/N018680/1).
Deutsche Forschungsgemeinschaft (SPP 1840, 152/6-2, SM 292/5-2).
EU Horizon 2020 Marie Skłodowska-Curie grant agreement No 101029393.
Agencia Estatal de Investigación (“Severo Ochoa” Center of Excellence CEX2019-000910-S, National Plan FIDEUA PID2019-106901GB-I00/10.13039 / 501100011033, FPI). Fundació Privada Cellex. Fundació Mir-Puig. Generalitat de Catalunya (AGAUR Grant No. 2017 SGR 1341, CERCA program).




\bibliography{Bibliography}
\bibliographystyle{ieeetr}

\end{document}